\documentclass{article}
\usepackage{fortschritte}
\usepackage{amssymb}
\def\bra{\langle}
\def\ket{\rangle}
\def\beq{\begin{equation}}                     %
\def\eeq{\end{equation}}                       %
\def\bea{\begin{eqnarray}}                     
\def\eea{\end{eqnarray}}                       
\begin {document}                 
\def\titleline{
%
%
%
Operator Product Expansion and\\                
Zero Mode Structure in logarithmic CFT          
}
\def\email_speaker{
{\tt 
%
%
flohr@itp.uni-hannover.de                       
}}
\def\authors{
%
%
%
%
%
Michael Flohr,\1ad\2ad$\!\!\!\!$\sp             
\phantom{mmmmmmmmmm}                            
Marco Krohn\2ad                                 
\phantom{m}                                     
}
\def\addresses{
%
%
%
%
                                               %
\begin{tabular}{cc}                            %
\1ad Physikalisches Institut &                 %
\2ad Institute for Theoretical Physics\\       %
University of Bonn &                           %
University of Hannover\\                       %
Nussallee 12, D-53115 Bonn, Germany &          %
Appelstr. 2, D-30167 Hannover, Germany         %
\end{tabular}                                  
                                               %
                                               %
}
\def\abstracttext{
%
%
The generic structure of 1-, 2- and 3-point    
functions of fields residing in indecomposable 
representations of arbitrary rank are given.   
These in turn determine the structure of the   
operator product                               %
expansion in logarithmic conformal field       %
theory. The crucial role of zero modes is      %
discussed in some detail.                      %
}
\large
\makefront
\section{Introduction}
During the last few years, logarithmic conformal field
theory (LCFT) has been established as a well-defined variety of
conformal field theories in two dimensions. The concept was considered in
its own right first by Gurarie \cite{Gurarie:1993xq}, Since then, a large
amount of work has appeared, see the reviews \cite{Flohr:2001zs,
Gaberdiel:2001tr} and references therein. The defining feature of a LCFT
is the occurence of indecomposable representations which, in turn, may lead to
logarithmically diverging correlation functions. Thus, in the standard example
of a LCFT a primary field $\phi(z)$ of conformal weight $h$ has a so-called 
logarithmic partner field $\psi$ with the characteristic properties
\begin{equation}\label{eq:1}
	\bra \phi(z)\phi(0)\ket = 0\,,\ \ \ \
	\bra \phi(z)\psi(0)\ket= Az^{-2h}\,,\ \ \ \
	\bra \psi(z)\psi(0)\ket=z^{-2h}\left(B - 2A\log(z)\right)\,.
\end{equation}
To this corresponds the fact that the highest weigth state $|h\ket$ 
associated to the primary field $\phi$ is the ground state of an irreducible 
representation which, however, is part of a larger, indecomposable, 
representation created from $|\tilde h\ket$, the state associated to $\psi$. 
The conformal weight is the eigenvalue
under the action of $L_0$, the zero mode of the Virasoro algebra, which in
such LCFTs cannot be diagonalized. Instead, we have
\begin{equation}\label{eq:2}
	L_0|h\ket = h|h\ket\,,\ \ \ \ L_0|\tilde h\ket = h|\tilde h\ket
	+ |h\ket\,.
\end{equation}
Thus, the two states $|h\ket$ and $|\tilde h\ket$ span a Jordan cell of rank two
with respect to $L_0$.
As can be guessed from eq.~(\ref{eq:1}), there must exist a zero mode which
is responsible for the vanishing of the 2-pt function of the primary field.
Another characteristic fact in LCFT is the existence of at least one field,
which is a perfect primary field, but whose operator product expansion (OPE)
with itself produces a logarithmic field. Such fields $\mu$ are called
pre-logarithmic fields \cite{Kogan:1997fd}. This is important, since in
many cases, the pre-logarithmic fields arise naturally forcing us then to
include the logarithmic fiels as well into the operator algebra. Note that
this implies that the fusion product of two irreducible representations is
not necessarily completely reducible into irreducible representations.
In fact, we know today quite a few LCFTs, where precisely this is the case,
such as ghost systems \cite{Krohn:2002gh}, WZW models at level zero or at
fractional level such as $\widehat{SU(2)}_{-4/3}$ \cite{Gaberdiel:2001ny,
Kogan:2001nj}, WZW models of 
supergroups such as $GL(1,1)$ \cite{Rozansky:1992td} or certain supersymmetric 
$c=0$ theories such as $OSP(2n|2n)$ or $CP(n|n)$ \cite{Gurarie:1999yx,
Read:2001pz}. Finally, many LCFTs are 
generated from free anticommuting fields such as the symplectic fermions
\cite{Kausch:2000fu}. Such LCFTs have an interesting fermionic structure where
logarithms may also arise in correlation functions involving spin zero
anticommuting fields. This is in contrast to free bosons, which typically 
do not directly appear in the conformal field theory, but only in form of
derivatives and exponentials of themselves.

In these notes, we generalize LCFT to the case of Jordan cells of arbitrary
rank, but we will restrict ourselves to the Virasoro algebra as the chiral
symmetry algebra to keep things simple. With some mild assumptions, the
generic form of 1-, 2- and 3-pt functions can be given such that only the
structure constants remain as free parameters. From this, the general 
structure of the OPE as well as some sort of selection rules that a general 
correlation function may be non-zero, are derived. The crucial role of
zero modes, in particular in the case of a LCFT generated from fermionic
fields, is emphasized. The results presented here, together with proofs and
further details, can be found in \cite{Flohr:2001tj,Krohn:2002gh}.
The computation of 4-pt and higher-point functions in the LCFT case is,
unfortunatley, more complicated. The interested reader might consult
\cite{Flohr:1998ew,Flohr:2000mc} for some discussion on this issue.

\section{1-, 2- and 3-pt functions}
Let $r$ denote the rank of the Jordan cells we consider. One can show, that
in LCFTs with Jordan cells with respect to (at least) the $L_0$ mode, the
$h=0$ sector necessarily must carry such a Jordan cell structure. Furthermore,
its rank defines the maximal possible rank of all Jordan cells. Thus, without
loss of generality, we can assume that the rank of all Jordan cells is equal
to $r$, other cases can easily be obtained by setting certain structure
constants to zero. Each Jordan cell contains one proper highest weight state
giving rise to one proper irreducible subrepresentation. We will label this
state for a Jordan cell with conformal weight $h$ by $|h;0\ket$. We choose
a basis in the Jordan cell with states $|h;k\ket$, $k=0,\ldots,r-1$, such that
eq.~(\ref{eq:2}) is replaced by
\begin{equation}\label{eq:3}
	L_0|h;k\ket = h|h;k\ket + |h;k-1\ket\ \ \textrm{for}\ \ 
	k=1,\ldots,r-1\,,\ \ \ \
	L_0|h;0\ket = h|h;0\ket\,.
\end{equation}
The corresponding fields will be denoted $\Psi_{(h;k)}$. Although the OPE
of two primary fields might produce logarithmic fields, we will further assume,
that primary fields {\slshape which are members of Jordan cells\/} are proper
primaries in the sense that OPEs among them only yield again primaries.

As discussed by Rohsiepe \cite{Rohsiepe:1996qj}, the possible structures
of indecomposable representations with respect to the Virasoro algebra
are surprisingly rich. Besides the defining condition eq.~(\ref{eq:3}),
further conditions have to be employed to fix the structure. The simplest
case is defined via the additional requirement
\begin{equation}\label{eq:L0qp}
       L_{1}|h;k\ket = 0\,,\ \ \ \ 0\leq k<r\,.
\end{equation}
This condition means that all fields spanning the Jordan cell are
quasi-primary. It will be our starting point in the following.
Under these assumptions, as shown in \cite{Flohr:1997wm}, the action of the 
Virasoro modes receives an additional non-diagonal term.
The off-diagonal action is defined via
$\hat{\delta}_{h_i}\Psi_{(h_j;k_j)}(z) = \delta_{ij}\Psi_{(h_j;k_j-1)}(z)$ for
$k_j>0$ and $\hat{\delta}_{h_i}\Psi_{(h_j;0)}(z) = 0$. Thus,
\begin{equation}\label{eq:viir}
   L_n \bra\Psi^{}_{(h_1;k_1)}(z_1)\ldots\Psi^{}_{(h_n;k_n)}(z_n)\ket =
   \sum_iz_i^n\left[z_i\partial_i + (n+1)(h_i+\hat{\delta}_{h_i})\right]
   \bra\Psi^{}_{(h_1;k_1)}(z_1)\ldots\Psi^{}_{(h_n;k_n)}(z_n)\ket
\end{equation}
for $n\in\mathbb{Z}$.
Only the generators $L_{-1}$, $L_0$, and $L_1$ of the M\"obius
group admit globally valid conservation laws, which usually are expressed
in terms of the so-called conformal Ward identities
\begin{equation}\label{eq:ward} 0 = \left\{\begin{array}{rcl}
         L_{-1} G(z_1,\ldots z_n) & = & \sum_i\partial_i G(z_1,\ldots z_n)
         \,,\\[0.2cm]
         L_0 G(z_1,\ldots z_n) & = & \sum_i(z_i\partial_i + h_i +
	 \hat{\delta}_{h_i})G(z_1,\ldots z_n)\,,\\[0.2cm]
	 L_1 G(z_1,\ldots z_n) & = & \sum_i(z_i^2\partial_i
	 + 2z_i[h_i + \hat{\delta}_{h_i}]) G(z_1,\ldots z_n)\,,
    \end{array}\right.
\end{equation}
where $G(z_1,\ldots z_n)$ denotes an arbitrary $n$-point function
$\bra\Psi_{(h_1;k_1)}(z_1)\ldots\Psi_{(h_n;k_n)}(z_n)\ket$ of primary
fields and/or their logarithmic partner fields. Here, we already have
written down the Ward identities in the form valid for proper
Jordan cells in logarithmic conformal field theories. Note that these are
now inhomogeneous equations. In principle,
this is all one needs to compute the generic form of all $n$-pt functions,
$n\leq 3$ upto structure constants. Thus, using freedom of scaling the fields,
the 1-pt functions turn out to be
\begin{equation}\label{eq:1pt}
	\bra\Psi^{}_{(h;k)}\ket = \delta_{h,0}\delta_{k,r-1}\,.
\end{equation}
The 2-pt and 3-pt functions can be written in a rather compact form by noting 
that derivaties of $z^{h}$ with respect to $h$ yields a logarithm. The structure
constants depend on both, the conformal weights as well as the {\slshape
total level\/} within the Jordan cells. One obtains
\begin{equation}\label{eq:2pt}
	\bra\Psi^{}_{(h;k)}(z)\Psi^{}_{(h',k')}(z')\ket =
	\sum_{j=r-1}^{k+k'}D_{(h;j)}\,\delta_{h,h'}
	\sum_{{0\leq i\leq k,0\leq i'\leq k'\atop
	i+i'=k+k'-j}}\frac{1}{i!i'!}\,(\partial_h)^i(\partial_{h'})^{i'}
	(z-z')^{-h-h'}
\end{equation}
for the 2-pt functions, and for the 3-pt functions analogously
\begin{eqnarray}\label{eq:3pt}
	\bra\Psi^{}_{(h_1;k_1)}(z_1)\Psi^{}_{(h_2,k_2)}(z_2)\Psi^{}_{(h_3;k_3)}
	(z_3)\ket &=&
	\sum_{j=r-1}^{k_1+k_2+k_3}C_{(h_1,h_2,h_3;j)}\!\!
	\sum_{{0\leq i_l\leq k_l,l=1,2,3\atop i_1+i_2+i_3=k_1+k_2+k_3-j}} 
	\!\!\frac{1}{i_1!i_2!i_3!}\nonumber\\ \times\,
	(\partial_{h_1})^{i_1}(\partial_{h_2})^{i_2}
	(\partial_{h_3})^{i_3}
	& & \!\!\!\!\!\!\!\!\!\!\!\!\!
	(z_{12})^{h_3-h_1-h_2}(z_{13})^{h_2-h_1-h_3}
	(z_{23})^{h_1-h_2-h_3}\,.
\end{eqnarray}

\section{OPEs}
It is now a simple matter to write down the generic form of OPEs. In essence,
we have to raise one index of the 3-pt structure constants with the help of
the inverse of the 2-pt structure constants, {\slshape i.e.}~the propagators.
Now, in the LCFT case, we have matrices instead, namely
\begin{equation}
	\left(D_{h,h'}\right)_{k,k'} \equiv \delta_{h,h'}
	\bra\Psi^{}_{(h;k)}(z_2)\Psi^{}_{(h',k')}(z_3)\ket\,,
\end{equation}
which is an upper triangular matrix and thus invertible, and
\begin{equation}
	\left(C_{(h_1;k1),h_2,h_3}\right)_{k_2,k_3} \equiv
	\bra\Psi^{}_{(h_1;k_1)}(z_1)\Psi^{}_{(h_2,k_2)}(z_2)\Psi^{}_{(h_3;k_3)}
	(z_3)\ket\,.
\end{equation}
The OPE is then given by the expression
\begin{equation}
	\Psi^{}_{(h_1;k_1)}(z_1)\Psi^{}_{(h_2;k_2)}(z_2) = 
	\lim_{z_1\rightarrow z_2}
	\sum_{(h_3;k_3)}\sum_k\left(C_{(h_1;k_1),h_2,h_3}\right)_{k_2,k}
	\left((D_{h_3,h_3})^{-1}\right)^{k,k_3}\Psi^{}_{(h_3;k_3)}(z_2)\,,
\end{equation}
where the limit means that we have to replace $z_{13}$ in the result by
$z_{23}$ which, in fact, will cancel all dependency on $z_3$. In this form,
the OPE does not obey a bound such as $k_3 \leq k_1+k_2$ for the so-called
J-levels within the Jordan blocks. For example, pre-logarithmic fields are
good primary fields, such that $k_1=k_2=0$, while there appears a term with 
$k_3=1$ on the right hand side. 

A better bound is given by the {\slshape zero mode content\/} of the fields.
This means the following: The basic fields of the conformal field theory 
might contain a certain number of zero modes
$\theta_0^{(\alpha)}$ such that $\bra 0|\theta_0^{(\alpha)} = 
\theta_0^{(\alpha)}|0\ket = 0$. These modes will come with canonical
conjugate modes $\xi_{(\alpha)}$, which are creators to the right as well as
to the left. Thus, the zero mode content $Z_0(\Psi)$ of a field $\Psi$ is 
defined as the total number of $\xi_{(\alpha)}$ modes in its mode expansion,
expressed in the modes of the basic fields. If the basic fields are
anticommuting fermions, we will have anticommuting pairs $\xi^\pm_{(\alpha)}$
instead such that we can define zero mode contents $Z_+(\Psi)$ and
$Z_-(\Psi)$ separately with $Z_0 = Z_++Z_-$. Explicitly known examples
of LCFTs do have realizations in fermionic free fields, and it turns out
that the definition above can be extended to pre-logarithmic fields in a 
consistent way by assigning them fractional values $Z_+$ and $Z_-$ such that
always $Z_0\in\mathbb{Z}$. In fact, a large class of LCFTs can be constructed
from ordinary conformal field theories by introducing additional zero modes
accompanied with a suitable deformation of the Virasoro modes, see
\cite{Fjelstad:2002ei} for details. The zero mode content does now provide
a bound for OPEs, namely
\begin{equation}
	Z_0(\Psi_{(h_3;k_3)})\leq Z_0(\Psi_{(h_1;k_1)}) + Z_0(\Psi_{(h_2;k_2)})
	\,.
\end{equation}

One of the best known examples for a LCFT is the $c=-2$ ghost system,
written in terms of two anticommuting spin zero fields $\theta^\pm(z)$.
The mode expansion reads
\begin{equation}
	\theta^\pm(z) = \xi^\pm+\theta_0^\pm\log(z)+\sum_{n\neq 0}
	\theta_n^\pm z^{-n}\,,
\end{equation}
where the modes $\xi^\pm$ are the creator zero modes, while the modes
$\theta^\pm_0$ are the annihilator zero modes, satisfying $\{\xi^\pm_{},
\theta^\mp_0\}=1$, $\{\theta^+_n,\theta^-_m\}=\frac{1}{n}\delta_{n+m,0}$. 
We get back the original $bc$ ghost system by setting
$c(z)=\left.\theta^-(z)\right|_{\theta^-_0=0}$ and $b(z)=\partial_z
\theta^+(z)$. Thus, the pair $\theta^-_0$ and $\xi^+$ of zero modes
is absent in the $bc$ system, and so is the logarithmic partner of
the identity field, $\Psi_{(h=0;1)}(z)=\mbox{:$\theta^+\theta^-$:}(z)$ with
state $|0;1\ket=\xi^+\xi^-|0;0\ket$, where $|0;0\ket = |0\ket$. Thus,
$Z_0(\Psi_{(0;0)}) = 2$, $Z_\pm(\Psi_{(0;0)})=1$, while the basic fermionic
fields obey $Z_0(\theta^\pm)=1$, $Z_\pm(\theta^\pm)=1$ and
$Z_\mp(\theta^\pm)=0$. This theory possesses a pre-logarithmic field $\mu$ of
conformal weight $h=-1/8$ with  
OPE $\mu(z)\mu(0)\sim z^{1/4}\left(\Psi_{(0;1)}(0) 
-2\log(z)\Psi_{(0;0)}(0)\right)A + z^{1/4}\Psi_{(0;0)}(0)B$. 
To make everything consistent, one assigns $Z_\pm(\mu)=1/2$. We mention for
completeness, that the excited twist field $\sigma$ with conformal weight
$h=3/8$ has to be assigned the values $Z_+(\sigma)=3/2$ and $Z_-(\sigma)=-1/2$
or vice versa.

\section{Zero mode content}
Let us briefly consider a much less trivial example, the ghost system
with $c=-26$, made out of a pair of anticommuting fiels of spin $2$ and $-1$,
respectivelyi \cite{Krohn:2002gh}. 
In general, the $(j,1-j)$ ghost system possesses $2j-1$
zero modes $b_{j-1},b_{j-2},\ldots,b_{1-j}$. The stress energy tensor reads
\begin{equation}
	T_{bc}=-j\mbox{:$b(\partial c)$:} + (1-j)\mbox{:$(\partial b)c$:}\,.
\end{equation}
Using a slight generalization of the deformation technique of 
\cite{Fjelstad:2002ei}, additional zero modes can be introduced by a
modification, shown here for the $j=2$ case,
\begin{equation}
	T_{\log}(z) = T_{bc}(z) + A\theta^-_1\partial b(z) + 
	B\theta^-_0z^{-1}\partial(z^2 b(z)) + A\theta^-_{-1}z^{-2}
	\partial(z^4 b(z))\,.
\end{equation}
These additional zero modes can be thought of as modes of $h\!=\!-1$ fields
$\theta^\pm$ with expansion
\begin{equation}
	\theta^\pm(z)=\xi^\pm_{-1}z^2+\xi^\pm_0 z + \xi^\pm_{+1}
	+\theta^\pm_{-1}\frac{z^2}{2}(\log(z)-\frac{3}{2})
	+\theta^\pm_0 z(\log(z)-1) + \theta^\pm_{+1}\log(z)
	+\sum_{|n|>1}\theta^\pm_n \frac{z^{-n+1}}{1-n}\,.
\end{equation}
Again, $b(z)=\partial^{2j-1}\theta^+(z)$ and $c(z)=\left.
\theta^-(z)\right|_{\theta^-_{j-1}=\theta^-_{j-2}=\ldots=\theta^-_{1-j}=0}$
such that the $\theta^\pm$ fields have twice as many zero modes as the
original $bc$ system, and $\{\xi^\pm_i,\theta^\mp_{-i}\}=\pm(-1)^{i+1}$. 
Although the modes of the modified stress energy tensor 
satisfy the Virasoro algebra, they do not act consistently on the space of
states, {\slshape e.g.}~$L_0|\xi^+_{-1}\ket=0$.

However, as explained in \cite{Krohn:2002gh}, the doubling of the zero modes
is not completely articifial, but does naturally imply that the conformal
field theory (CFT) now lives on a hyperellptic Riemann surface, viewed as a 
double covering of the complex plane or Riemann sphere. Thus, we actually have
a CFT on each of the sheets, such that the full CFT is the tensor product of
the individual ones with $T_{\log}=T^{(1)}_{\log}+T^{(2)}_{\log}$ such that
$[T^{(1)}_{\log},T^{(2)}_{\log}]=0$. In fact, this is possible and yields a
consistent CFT provided we identify the zero modes on the different sheets with
each other as $\theta_i^{(1),\pm}=-\theta_i^{(2),\mp}$ and
$\xi_i^{(1),\pm}=\xi_i^{(2),\mp}$ for $i=-1,0,1$. This yields the Virasoro
algebra for $T_{\log}$ with total central charge $c=-52$ and a correct action
of its modes on the space of states. The resulting theory possesses 
indecomposable representations despite the fact that the action of $L_0$
remains diagonal. The construction generalizes to other
ghost systems, but is is not yet clear, how the construction works for
higher ramified covering with more than two sheets. 

This shows that zero modes are at the heart of LCFTs. Furthermore,
the zero mode content provides strong conditions on whether correlation
functions can actually be non-zero. It appears that fields $\Psi_{(h;k)}$
forming Jordan blocks have a well defined {\slshape even\/} 
zero mode content $Z_0=Z_++Z_-$ 
with $Z_+=Z_-$. Fermionic fields, in turn, are characterized by $Z_+\neq Z_-$ 
but still satisfy $Z_\pm\in\mathbb{Z}$. These fields are denoted by
$\Theta_{(h;k^+,k^+)}$. There are no examples known where such fields do
form indecomposable structures, but this is only due to the fact that no
LCFTs with a sufficiently high number of genuine zero modes have been 
explicitly examined yet. Finally, pre-logarithmic, or more generally, twist
fields, have fractional zero mode contents, $Z_\pm\in\mathbb{Q}-\mathbb{Z}$,
and are denoted by $\mu_\alpha$. Such fields are generally believed to reside
in irreducible representations. A generic correlation function is then
of the form
\begin{equation}
	G=\left\langle
	\prod_i\Psi_{(h_i;k_i)}(z_i)\prod_j\Theta_{(h_j;k^+_k,k^-_j)}(w_j)
	\prod_l\mu_{\alpha_l}(u_l)\right\rangle\equiv
	\bra\prod_i\Psi_i\prod_j\Theta_j\prod_l\mu_l\ket\,.
\end{equation}
The zero mode contents implies now that $G=0$ unless all three conditions
\begin{eqnarray}\label{cond1}
	& & \mathbb{Z}_+\ \ni\ Z_0(G) =
	\sum_iZ_0(\Psi_i)+\sum_jZ_0(\Theta_j)+\sum_lZ_0(\mu_l) 
	\geq 2(r-1)\,,\\ \label{cond2}
	& &\sum_j Z_+(\Theta_j) = \sum_j Z_-(\Theta_j)\ \in \mathbb{Z}\,,\\
	\label{cond3} & &
	\sum_l Z_+(\mu_l)\ \in\ \mathbb{Z}\ \ \ \textrm{and}\ \ \
	\sum_l Z_-(\mu_l)\ \in\ \mathbb{Z}
\end{eqnarray}
are satisfied. These are very powerfull statements since they imply further
that we can relax our condition that the logarithmic partners have to be
quasi-primary. In fact, the action of $L_n$, $n=-1,0,1$, in the Ward identities 
eq.~(\ref{eq:ward}) will yield new correlation functions, $L_n G=\sum_kG_k'$.
If there are contributions from fields failing to be quasi-primary, then
these can be neglected if the resulting correlation functions $G'$ do not 
anylonger satisfy eqs (\ref{cond1}--\ref{cond3}). Since $L_n$ act as 
derivations, only one field in the correlator is modified in each term. Thus,
if the zero mode content of a non-quasi-primary term differs from the
original zero mode content such that the balance is broken, it will not
contribute to the correlation function $G$ since it does not affect the
Ward identities. Hence, $L_n$ implements a BRST like structure on the complex 
spanned by $Z_+$ and $Z_-$, as anticipated in \cite{Moghimi-Araghi:2002gk}.

In summary, we have provided the general structure of correlation functions
and OPEs for LCFT with arbitrary high rank Jordan cells. We found strong
constraints for correlation functions to be non-zero, intimately linked to
the zero mode content of the involved fields. It remains an open
problem, however, what the modular properties of such higher rank LCFTs
are. These are only knonw in the rank two case \cite{Flohr:1995ea}.

{\bf Acknowledgement} The research of M.\,F.\ is supported by
the European Union network HPRN-CT-2002-00325 and the research of M.\,F.\ and
M.\,K.\ is supported by the string theory network
(SPP no.\ 1096), Fl 259/2-2, of the Deutsche Forschungsgemeinschaft.


\begin{thebibliography}{77}
\bibitem{Fjelstad:2002ei}
	J.~Fjelstad, J.~Fuchs, S.~Hwang, A.~M.~Semikhatov and I.~Y.~Tipunin,
	Nucl.\ Phys.\ B {\bf 633} (2002) 379
	[arXiv:hep-th/0201091].
\bibitem{Flohr:1995ea}
	M.~A.~I.~Flohr,
	Int.\ J.\ Mod.\ Phys.\ A {\bf 11} (1996) 4147
	[arXiv:hep-th/9509166].
\bibitem{Flohr:1996vc}
	M.~A.~I.~Flohr,
	Int.\ J.\ Mod.\ Phys.\ A {\bf 12} (1997) 1943
	[arXiv:hep-th/9605151].
\bibitem{Flohr:1997wm}
	M.~A.~I.~Flohr,
	Nucl.\ Phys.\ B {\bf 514} (1998) 523
	[arXiv:hep-th/9707090].
\bibitem{Flohr:1998ew}
	M.~A.~I.~Flohr,
	Phys.\ Lett.\ B {\bf 444} (1998) 179
	[arXiv:hep-th/9808169].
\bibitem{Flohr:2000mc}
	M.~A.~I.~Flohr,
	arXiv:hep-th/0009137.
\bibitem{Flohr:2001tj}
	M.~A.~I.~Flohr,
	Nucl.\ Phys.\ B {\bf 634} (2002) 511
	[arXiv:hep-th/0107242].
\bibitem{Flohr:2001zs}
	M.~A.~I.~Flohr,
	Int.\ J.\ Mod.\ Phys.\ A {\bf 18} (2003) 4497
	[arXiv:hep-th/0111228].
\bibitem{Gaberdiel:2001ny}
	M.~R.~Gaberdiel,
	Nucl.\ Phys.\ B {\bf 618} (2001) 407
	[arXiv:hep-th/0105046].
\bibitem{Gaberdiel:2001tr}
	M.~R.~Gaberdiel,
	Int.\ J.\ Mod.\ Phys.\ A {\bf 18} (2003) 4593
	[arXiv:hep-th/0111260].
\bibitem{Gurarie:1993xq}
	V.~Gurarie,
	Nucl.\ Phys.\ B {\bf 410} (1993) 535
	[arXiv:hep-th/9303160].
\bibitem{Gurarie:1999yx}
	V.~Gurarie and A.~W.~W.~Ludwig,
	J.\ Phys.\ A {\bf 35} (2002) L377
	[arXiv:cond-mat/9911392].
\bibitem{Kausch:2000fu}
	H.~G.~Kausch,
	Nucl.\ Phys.\ B {\bf 583} (2000) 513
	[arXiv:hep-th/0003029].
\bibitem{Kogan:1997fd}
	I.~I.~Kogan and A.~Lewis,
	Nucl.\ Phys.\ B {\bf 509} (1998) 687
	[arXiv:hep-th/9705240].
\bibitem{Kogan:2001nj}
	I.~I.~Kogan and A.~Nichols,
	Int.\ J.\ Mod.\ Phys.\ A {\bf 17} (2002) 2615
	[arXiv:hep-th/0107160].
\bibitem{Krohn:2002gh}
	M.~Krohn and M.~A.~I.~Flohr,
	JHEP {\bf 0301} (2003) 020
	[arXiv:hep-th/0212016].
\bibitem{Moghimi-Araghi:2002gk}
        S.~Moghimi-Araghi, S.~Rouhani and M.~Saadat,
        Int.\ J.\ Mod.\ Phys.\ A {\bf 18} (2003) 4747
        [arXiv:hep-th/0201099].
\bibitem{Read:2001pz}
	N.~Read and H.~Saleur,
	Nucl.\ Phys.\ B {\bf 613} (2001) 409
	[arXiv:hep-th/0106124].
\bibitem{Rohsiepe:1996qj}
	F.~Rohsiepe,
	arXiv:hep-th/9611160.
\bibitem{Rozansky:1992td}
	L.~Rozansky and H.~Saleur,
	Nucl.\ Phys.\ B {\bf 389} (1993) 365
	[arXiv:hep-th/9203069].
\end{thebibliography}
\end{document}